\documentstyle[seceq,epsbox,preprint]{jpsj}

\def\runtitle
{Scaling Law and Aging Phenomena in the Random Energy Model}
\def\runauthor{Munetaka {\sc Sasaki} and Koji {\sc Nemoto}}
\title{Scaling Law and Aging Phenomena in the Random Energy Model}
\author{Munetaka {\sc Sasaki} and Koji {\sc Nemoto}}

\inst{Division of Physics, Hokkaido University, Sapporo 060-0810}
\recdate{\phantom{October 4, 1999}}
\abst{We study the effect of temperature shift on aging phenomena in the 
Random Energy Model (REM). From calculation on the correlation function and 
simulation on the Zero-Field-Cooled magnetization, we find that the REM 
satisfies a scaling relation even if temperature is shifted. Furthermore, this 
scaling property naturally leads to results obtained in experiment 
and the droplet theory. 
}
\kword{aging,scaling,random energy model,zero-field-cooled magnetization,
droplet theory
}
\begin{document}
\sloppy
\maketitle
\section{Introduction}\label{sec:introduction}
The aging, which are dynamical behaviors largely depending on the 
history of system, is one of the most striking phenomena in the complex 
systems such as spin glasses, glasses, polymers and proteins, and 
has been studied vigorously from both theoretical and experimental aspects. 
In experiments on spin glasses, one of the most familiar methods for the 
study of the aging phenomena is the measurement of the Zero-Field-Cooled 
(ZFC) magnetization\cite{ZFC1,ZFC2}. 
In this measurement, the sample is quenched to a temperature \( T \) 
below \( T_{\rm c} \) in zero field. After a waiting time \( t_{\rm w} \), 
a weak magnetic field is applied and the magnetization \( M(t)\) 
is recorded as a 
function of observation time \( t \). In this experiment, 
the length of \( t_{\rm w} \) determines the degree of equilibration 
and aging appears as the \( t_{\rm w} \) dependence, especially, 
the peak of the relaxation rate \( S(t) \equiv \frac{\partial M(t)}
{\partial t} \) emerges around \( t_{\rm w} \) as if the system remembers 
how the equilibration has proceeded before the observation. 

Granberg {\it et al.}\cite{ZFCTS} modified this experiment by shifting the 
waiting temperature from \( T \) to \( T-\Delta T \) and observed that, 
although \( t_{\rm w}\) was unchanged, the maximum of \( S(t) \) 
shifted to left with increasing \( \Delta T \) as if \( t_{\rm w} \)
decreased. We refer to the position of this 
maximum as apparent wait time \( t_{\rm w}^{\rm app} \). This suggests 
that the equilibration at \( T-\Delta T \) during \( t_{\rm w} \) 
only corresponds to the one at \( T \) during \( t_{\rm w}^{\rm app} \). 
For \( \Delta T < 0 \), the position shifts to right. 
They also investigated the relation between \( t_{\rm w} \) and 
\( \Delta T \) and found
\begin{equation}
\log(t_{\rm w}^{\rm app}/t_{\rm w}) \approx 
-\alpha(t_{\rm w}) \Delta T,
\label{eqn:relation1}
\end{equation}
with the result that the coefficient \( \alpha(t_{\rm w}) \) is a monotonically increasing function. 

From the theoretical aspects, this phenomenon is demonstrated by the droplet 
theory very well\cite{droplet1}, and is reproduced in the study of the two and 
three dimensional EA ising spin glass model\cite{MCS} and a hierarchical 
diffusion model\cite{hierarchical}. In this manuscript, we study this 
phenomenon with the Random Energy Model (REM)\cite{Bouchaud,REM2,REM3}. 
At first, we calculate a correlation function with the temperature shift 
and confirm that this correlation function satisfies a scaling law. 
Next, we carry out the simulation to observe the temperature shift ZFC 
magnetization. As a result, we confirm that this magnetization also satisfies 
the same scaling law as the correlation function, which naturally leads to 
the relation eq.~(\ref{eqn:relation1}). 

The organization of this manuscript is following. In section 2, we introduce 
the REM. In section 3, we define the temperature shift correlation function 
and calculate it. In section 4, we observe the temperature shift ZFC 
magnetization by simulation and analyze the mechanism of this phenomenon. 
In section 6, we summarize this paper with some discussions. 

\section{The Random Energy Model}
The REM is schematically shown in Fig.~1\cite{Bouchaud}. The bottom points represent the 
accessible states of this system. The length 
of each branch represents the barrier energy \( E \), over which the system 
goes from one state to another. This energy is an independent random variable 
distributed as
\begin{equation}
\rho (E) {\rm d}E = \frac{{\rm d} E}{T_{\rm c}}\exp[-E/T_{\rm c} ],
\label{eqn:simplerho}
\end{equation}
where \( T_{\rm c} \) is the transition temperature. 

From the Arrhenius law, 
the relaxation time \( \tau(\alpha:T) \), the average time
 for the system to escape from the state \( \alpha \) at \( T \), 
is related to \( E(\alpha) \) as
\[ \tau(\alpha:T) = \tau_0 \exp[ E(\alpha)/T], \]
where \( \tau_0 \) is a microscopic time scale. Hereafter \( \tau_0 \) 
is used as the unit of time and set to \( 1 \). 
From 
eq.~(\ref{eqn:simplerho}), the distribution of \( \tau \) can be written as
\begin{equation}
p_x(\tau) {\rm d} \tau = \frac {x }{\tau^{x+1}}{\rm d} \tau 
\hspace{1cm}(\tau \geq 1),
\label{eqn:sinpleptau}
\end{equation}
where \( x \equiv T/T_{\rm c} \).
From eq.~(\ref{eqn:sinpleptau}), 
it is easily shown that the averaged relaxation time \(\langle \tau \rangle \) 
is \( x/(x-1) \) for \( x>1 \) and infinite for \( x \le 1 \). 
This means that the transition from the ergodic phase to the non-ergodic phase 
occurs at \( T_{\rm c} \). 

For dynamics, we employ the following simple Markoff process. 
First, the system is thermally activated 
from \( \alpha \) in unit time with the probability
\begin{equation}
W(\alpha:T)=\exp[-E(\alpha)/T].
\label{eqn:defofW}
\end{equation}
Since the time evolution is Markoffian, we can easily evaluate the probability 
\( q(\alpha,t:T) {\rm d}t \) that 
for the event to occur during \( t+t_0 \) and \( t+t_0+{\rm d}t \) with 
knowing the system is in the state \( \alpha \) at \( t_0 \) as
\begin{equation}
q(\alpha,t:T) {\rm d}t=W(\alpha:T)\exp[-W(\alpha:T)t]{\rm d}t.
\label{eqn:defofq}
\end{equation}
We generate an event time \( t \) according to \( q(\alpha,t:T) \) to perform 
event-driven Monte Carlo simulation. When the magnetic field \( H \) 
is applied, 
we replace \( E(\alpha) \) in eq.~(\ref{eqn:defofW}) with \( E+HM_{\alpha} \), 
where \( M_{\alpha} \) is the magnetization of the state \( \alpha \). 
The magnetization is chosen randomly and independently from a distribution 
\( D(M) \) with zero mean. In this simulation, we choose the uniform 
distribution between \( -1 \) and \( +1 \) for \( D(M) \). 

After the activation, the system falls to one of all states with equal 
probability. In the limit 
\(N \rightarrow \infty\), we can neglect the possibility that 
the system has stayed one state more than twice in our finite time simulation. 
Therefore, in this simulation, we create a new state by using the distribution 
functions \( \rho(E) \) and \( D(M) \) whenever activation occurs. 
\section{The Temperature Shift Correlation Function}
At first, we define the correlation function mentioned in 
\S\ref{sec:introduction}. 
This correlation function is observed in the following procedure. 
We consider the infinitely high temperature limit for the initial 
condition. In other words, the initial state is chosen randomly at \( t=0 \). 
Then, the system is kept at \( T-\Delta T \) during \( t_{\rm w} \) and 
heated (or cooled) to \( T \) after that. Here, the magnetic field \( H \) is always zero. 
In this procedure, we observe the correlation of magnetization 
between the time \( t_{\rm w} \) and \( t_{\rm w}+t \), which we 
hereafter refer to as \( C_{\Delta T}(t,t_{\rm w}) \). It is explicitly 
written as
\begin{equation}
C_{\Delta T}(t,t_{\rm w}) = \sum_{\alpha,\beta} M_{\beta}
G_{\beta\alpha}(t)M_{\alpha} P_{\alpha}(t_{\rm w}),
\label{eqn:CF1}
\end{equation} 
where \( P_{\alpha}(t_{\rm w}) \) is the probability that the system 
is found at state \( \alpha \) at \( t_{\rm w} \) and 
\( G_{\beta\alpha}(t) \) is the probability that the system which 
initially stays at \( \alpha \) reaches \( \beta \) at \(t\). 
Because \( M_{\alpha} \) is 
independent random variable with zero mean, eq.~(\ref{eqn:CF1}) is reduced 
to the autocorrelation function by taking the average over the distribution of 
the magnetization: 
\begin{equation}
C_{\Delta T}(t,t_{\rm w}) = \overline{M^2}\sum_{\alpha} 
G_{\alpha\alpha}(t)P_{\alpha}(t_{\rm w}),
\label{eqn:CF2}
\end{equation} 
where \( \overline{M^2} \) is the variance of the distribution \( D(M) \). 
Finally, we ignore the possibility that the system is activated 
from \( \alpha \) 
between \( t_{\rm w} \) and \( t+t_{\rm w} \) and we still find the system 
at \( \alpha \) at time \( t+t_{\rm w} \) in the limit 
\( N\rightarrow \infty \). As the result, eq.~(\ref{eqn:CF2}) is further 
reduced to 
\begin{equation}
C_{\Delta T}(t,t_{\rm w}) =\overline{M^2}\sum_{\alpha} 
\exp[-W(\alpha:T)t]P_{\alpha}(t_{\rm w}).
\label{eqn:CF3}
\end{equation} 

Bouchaud and Dean\cite{Bouchaud} calculated the Laplace transformation of 
\( P_{\alpha}(t_{\rm w}) \) and obtain the result, 
\begin{equation}
{\hat P}_{\alpha}(E)=\frac{E^{x_1}}{Nx_1c(x_1)}\frac{\tau(\alpha:T-\Delta T)}
{\{E\tau(\alpha:T-\Delta T)+1\}},
\label{eqn:hatPalpha}
\end{equation}
where \( x_1 \equiv \frac{T-\Delta T}{T_{\rm c}} \) and
\begin{equation}
c(x)\equiv\Gamma(x)\Gamma(1-x).
\end{equation}
The inverse Laplace transformation of eq.~(\ref{eqn:hatPalpha}) leads us to 
\begin{equation}
P_{\alpha}(t_{\rm w})=\frac{1}{Nx_1c(x_1)\Gamma(x_1)}
\int_0^{t_{\rm w}} {\rm d}s s^{x_1-1}\exp\left[-\frac{t_{\rm w}-s}
{\tau(\alpha:T-\Delta T)}\right].
\label{eqn:solP}
\end{equation}

Now, let us rewrite \( W(\alpha:T) \) as
\begin{equation}
W(\alpha:T)=\tau(\alpha:T-\Delta T)^{-\gamma},
\end{equation}
where \( \gamma=\frac{T-\Delta T}{T} \). Substituting this equation and 
eq.~(\ref{eqn:solP}) into eq.~(\ref{eqn:CF3}), we obtain
\begin{equation}
C_{\Delta T}(t,t_{\rm w})=\frac{\overline{M^2}}{c(x_1)\Gamma(x_1)}
\int_{1}^{\infty}{\rm d}\tau \tau^{-x_1-1}
\exp[-\tau^{-\gamma}t]
\int_0^{t_{\rm w}}{\rm d}s s^{x_1-1}\exp\left[-\frac{t_{\rm w}-s}{\tau}\right].
\end{equation}
By changing the variables as \( u\equiv t_{\rm w}/\tau, v\equiv s/\tau \) 
and replacing the upper limit of $u$-integral with \( \infty \) in the 
assumption \( t_{\rm w} \gg 1 \), we find the following scaling relation, 
\begin{equation}
C_{\Delta T}(t,t_{\rm w})={\tilde C}_{\Delta T}(t/t_{\rm S}),
\end{equation}
where 
\begin{equation}
t_{\rm S}=t_{\rm w}^{\gamma},
\end{equation}
and
\begin{equation}
{\tilde C}_{\Delta T}(X)=\frac{\overline{M^2}}{c(x_1)\Gamma(x_1)}
\int_0^{\infty} \frac{{\rm d} u}{u}\exp[-Xu^{\gamma}-u]\int_0^{u}
\frac{{\rm d} v}{v} v^{x_1} {\rm e}^v.
\end{equation}
We can easily check that for the case \( \Delta T = 0 \), 
this relation is reduced to the \( t/t_{\rm w} \) scaling as shown in 
ref.~\citen{Bouchaud}. 

Now, we are interested in the asymptotic behaviors of 
\( {\tilde C}_{\Delta T}(X) \) in the limit \( X\ll 1\) and \( X \gg 1 \). 
For \( X \gg 1 \), we obtain
\begin{subequations}
\begin{eqnarray}
\frac{C_{\Delta T}(t,t_{\rm w})}{\overline{M^2}}
&=&\frac{1}{\gamma c(x_1)\Gamma(x_1)}
\int_0^{\infty}\frac{{\rm d}s}{s}\exp[-s-(s/X)^{\frac{1}{\gamma}}]
\int_0^{(s/X)^{\frac{1}{\gamma}}}\frac{{\rm d}v}{v} v^{x_1}{\rm e}^v\nonumber\\
&\approx &\frac{1}{x_1 \gamma c(x_1)\Gamma(x_1)}
\int_0^{\infty}\frac{{\rm d}s}{s}{\rm e}^{-s}(s/X)^{x_1}\nonumber \\
&=& \frac{\Gamma(x_2)}{x_1 \gamma c(x_1)\Gamma(x_1)}X^{-x_2},
\label{eqn:asymptA}
\end{eqnarray}
where \( x_2 \equiv \frac{T}{T_{\rm c}} \). In the case \( X\ll 1 \) and \( x_1+\gamma > 1 \), we find
\begin{eqnarray}
\frac{C_{\Delta T}(t,t_{\rm w})}{\overline{M^2}} &=& 1-\frac{1}{\gamma c(x_1)\Gamma(x_1)}
\int_0^{\infty}\frac{{\rm d}s}{s}\exp[-(s/X)^{\frac{1}{\gamma}}]
\{ 1-{\rm e^{-s}} \}
\int_0^{(s/X)^{\frac{1}{\gamma}}}\frac{{\rm d}v}{v} v^{x_1}{\rm e}^v\nonumber\\
&\approx& 1-\frac{1}{\gamma c(x_1)\Gamma(x_1)}\int_0^{\infty}
\frac{{\rm d}s}{s}\{ 1-{\rm e^{-s}} \}(s/X)^{\frac{x_1-1}{\gamma}}\nonumber \\
&=& 1-\frac{\Gamma(\frac{x_1+\gamma-1}{\gamma})}{(1-x_1) c(x_1)\Gamma(x_1)}
X^{\frac{1-x_1}{\gamma}}.
\label{eqn:asymptB}
\end{eqnarray}
For \( X\ll 1 \) and \( x_1+\gamma < 1 \),
\begin{eqnarray}
\frac{C_{\Delta T}(t,t_{\rm w})}{\overline{M^2}} 
&=& 1-\frac{1}{c(x_1)\Gamma(x_1)}
\int_0^1\frac{{\rm d} s}{s}s^{x_1}\int_0^{\infty}\frac{{\rm d} u}{u}
u^{x_1}{\rm e}^{-u(1-s)} \{1-\exp[-Xu^{\gamma}]\} \nonumber\\
&\approx&1-\frac{X}{c(x_1)\Gamma(x_1)}\int_0^1\frac{{\rm d} s}{s}s^{x_1}
\int_0^{\infty}\frac{{\rm d} u}{u}u^{x_1+\gamma}{\rm e}^{-u(1-s)}\nonumber\\
&=&1-\frac{c(x_1+\gamma)}{c(x_1)} X.
\label{eqn:asymptC}
\end{eqnarray}
\end{subequations}

Next, we evaluate eq.~(\ref{eqn:CF3}) by simulation to check the
validity of these results. 
Random average is taken over \( 10^6 \) samples. In this simulation, 
we fix \( x_1 \) to \( 0.4 \) 
and change \( x_2 \) and \( t_{\rm w} \) as \( 0.2,0.4,0.6,0.8,1.2 \) 
and \( 10,10^2,10^3,10^4,10^5 \), respectively. 
In Fig.~2, the scaling 
plot for all \( x_2 \) is shown for (a) \(t/t_{\rm S} \gg 1 \) and 
(b) \( t/t_{\rm S}\ll 1 \). We can see that the scaling holds 
very well and the exponent in eqs.~(\ref{eqn:asymptA}),
  ~(\ref{eqn:asymptB}) and~(\ref{eqn:asymptC}) are rather correct, 
although a little deviation is observed on the coefficients 
for \( t/t_{\rm S}\ll 1 \). 

\section{The Result of the Temperature Shift ZFC Simulation}
In this section, we show the result of simulations on the temperature shift 
ZFC magnetization with \( T_{\rm c}=1.0 \) and \( T=0.5 \). 
The number of samples for random average is \( 5\times10^7 \). 
The amplitude of field is \( 0.1 \). 
We prepare the initial condition in the infinitely high temperature limit, 
the same as the temperature shift correlation function. 
During \( t_{\rm w} \), the temperature of the system is kept at 
\( T-\Delta T \). After that, the temperature is changed to \( T \) and 
the field \( H \) is applied. We performed the simulations for the 36 
different cases with \( t_{\rm w}=10^3,3\times10^3,10^4,3\times 10^4 \) and 
\( \Delta T = 0.08,0.06,\ldots,0,-0.02,\ldots,-0.08 \). 

In Fig.~3, we plot the relaxation rate \( S(t) \) for 
\( t_{\rm w}=3\times 10^4 \) and all \( \Delta T \). 
We can see the position of the peak shifts to left with increasing 
\( T-\Delta T \), although its shape becomes broader. 
Next, we checked whether the same scaling relation as 
\( C_{\Delta T}(t,t_{\rm w}) \) holds or not. In Fig.~4, we plot 
\( S(t) \) as a function of \( t/t_{\rm S} \) for all \( t_{\rm w} \) and 
\( \Delta T \). We can confirm the scaling law. 
Although the position of peak depends on \( \Delta T \) to some extent 
in our simulation, we consider that this dependence vanishes if 
we take \( \Delta T/T \) as small as that used in
experiment\cite{ZFCTS}. 
By neglecting this \( \Delta T \) 
dependence, we can roughly estimate the position of peak as 
\( t_{\rm w}^{\rm app}\approx t_{\rm S}=t_{\rm w}^{\frac{T-\Delta T}{T}} \) 
and obtain the relation eq.~(\ref{eqn:relation1}) with 
\( \alpha (t_{\rm w})= \frac{\log (t_{\rm w})}{T} \). This result is the same 
as the one from the droplet theory including the coefficient 
\( \alpha(t_{\rm w}) \)\cite{MCS}. 

In order to investigate how the equilibration proceeds during 
\( t_{\rm w} \), we examine the energy distribution \(P(E,t)(t\le t_{\rm w})\) 
which is defined as the probability density that the system is found 
at time \( t \) in one of the states whose energy is \( E \). In Fig.~5, 
we show how \( P(E,t) \) changes with time. At first sight, we can see 
that \( P(E,t) \) has 
a peak at some point \( E^* \), which shifts to right with increasing \( t \). 
The value of \( E^* \) is roughly determined how far the system 
can be activated during time interval \( t \), and estimated as
\begin{equation}
E^* \approx T \log t.
\label{eqn:shiftspeed}
\end{equation}
For \( E \le E^* \) the distribution is well equilibrated, so that 
the exponent \( \alpha_1 \) is given as
\begin{equation}
\alpha_1 = \frac1T -\frac{1}{T_{\rm c}}. 
\end{equation}
Note that this exponent changes its sign at \( T = T_{\rm c} \). 
While, the other part \( E > E^* \) is not equilibrated and 
the exponent \( \alpha_2 \) is equal to that of \( \rho(E) \), {\rm i.e.},
\begin{equation}
\alpha_2 = -\frac{1}{T_{\rm c}}.
\end{equation}

Equation~(\ref{eqn:shiftspeed}) suggests that the speed of the peak shift 
varies with the temperature. In Fig.~6, we show \( P(E,t) \) for 
\( t=10^4 \) and all \( \Delta T \). We can see that the position of 
the peak shifts to right just as Fig.~3. From these nature of 
\( P(E,t) \), we understand the relation eq.~(\ref{eqn:relation1}) 
as follows: During \( t_{\rm w} \) at \( T-\Delta T \), the peak shifts to 
\( E^* \approx (T-\Delta T)\log(t_{\rm w}) \). We can expect that, in this 
experiment, the magnetization most strongly changes at the time scale 
in which the system can be activated to \( E^* \) at \(T\). Therefore, 
we can estimate \( t_{\rm w}^{\rm app} \) from the condition
\begin{equation}
(T-\Delta T)\log(t_{\rm w}) \approx T \log(t_{\rm w}^{\rm app}),
\label{eqn:CDfortwapp}
\end{equation}
which leads to eq.~(\ref{eqn:relation1}) 
with \( \alpha (t_{\rm w})= \frac{\log (t_{\rm w})}{T} \), 
as mentioned above. 
\section{Summary and Discussions}
We have shown that the REM satisfies a scaling law even if the temperature is 
shifted and time \( t \) is scaled by 
\( t_{\rm S}=t_{\rm w}^{1-\frac{\Delta T}{T}} \). For the case 
\( \Delta T =0 \), \( t/t_{\rm w} \)-type scaling is reported in the 
experiment\cite{scaling1} and simulation on the three-dimensional 
\( \pm J \) Ising spin-glass model\cite{scaling2}, although some 
systematic deviations from the scaling remain on the experiment. It is interesting 
to check whether these systems satisfy the scaling like the REM or not. 

Next, we consider why the droplet theory leads to the same result 
as the REM. In the droplet theory, the crossover from equilibrium relaxation 
to non-equilibrium one occurs when the droplet created after the 
field application grows to the size of the droplet created before that, 
and we can expect that peak of \( S(t) \) emerges there. Therefore, 
\( t_{\rm w}^{\rm app} \) is determined by the condition
\label{eqn:DLcondition}
\begin{equation}
L(t_{\rm w},T-\Delta T)\approx L(t_{\rm w}^{\rm app},T),
\end{equation}
where \( L(t,T) \) denotes the characteristic size of the droplet 
after the age \( t \) at \( T \), and is given from the condition
\begin{equation}
B(T,L(t,T))=T\log(t),
\end{equation}
and \( B(T,L) \) denotes the barrier energy that the system has to 
overcome to create the droplet of size \( L \), and is given as 
\( B(T,L)=V(T) L^{\psi} \) explicitly. If \( T \) dependence of 
\( V(T) \) is week enough, the condition eq.~(\ref{eqn:DLcondition}) 
is reduced to eq.~(\ref{eqn:CDfortwapp}). 

Recently it was reported that \( t_{\rm w}^{\rm app} \) converges 
to \( 0 \) for somewhat large \( |\Delta T| \) (\( |\Delta T/T|>0.07 \) in 
this experiment) regardless of its sign\cite{ZFCTS2}. 
This suggests that the equilibration 
at \( T-\Delta T \) does affect the one at \( T \). In the droplet 
theory\cite{chaoticDP1,chaoticDP2}, 
it is considered that the {\it chaotic} change of the equilibrium spin 
configuration against the temperature variation causes this phenomenon. But 
the REM has no reason why this phenomenon is observed. Actually, 
this phenomenon is not observed in our simulation even in the region 
\( |\Delta T/T| \le 0.16 \). But we can show that 
this situation is improved by treating 
the Multi-layer Random Energy Model (MREM), 
the details of which will be presented elsewhere\cite{tobesubmitted}.
\section*{Acknowledgements}
We would like to thank H. Yoshino for fruitful discussions and
suggestions on the manuscript. The numerical calculations were performed 
on an Origin 2000 at Division of Physics, Graduate school of Science, 
Hokkaido University. 

\newpage
\noindent
{\bf \large FIGURE CAPTIONS}

\vspace*{3mm}\noindent

\vspace*{3mm}\noindent
Fig.~1  Structure of the Random Energy Model. 

\vspace*{3mm}\noindent
Fig.~2  (a) \( C_{\Delta T}(t,t_{\rm w}) \) and (b) 
        \( 1-C_{\Delta T}(t,t_{\rm w}) \) is plotted as a function of 
        \( t/t_{\rm S} \) for all \( t_{\rm w} \) and \( x_2 \). 
        Each line is the asymptotic behavior estimated in 
        eqs.~(\ref{eqn:asymptA}),~(\ref{eqn:asymptB}) and~(\ref{eqn:asymptC}).

\vspace*{3mm}\noindent
Fig.~3  \( S(t) \) vs. \( t \) for \( t_{\rm w}=3\times 10^4 \) and 
        \( \Delta T =0.08,0.06,\ldots,0,-0.02,\ldots,-0.08 \) 
        (from left to right). 

\vspace*{3mm}\noindent
Fig.~4  The scaling plot of \( S(t) \) for all \( t_{\rm w} \) and 
        \( \Delta T \). 

\vspace*{3mm}\noindent
Fig.~5  \( P(E,t) \) for \( T=0.5 \) at 
        \( t=10^{0.5},10,\ldots,10^{3.5} \) (from left to right).

\vspace*{3mm}\noindent
Fig.~6  \( P(E,t) \) for \( t=10^4 \) and \( \Delta T= 0.08,0.06,\ldots,0,
        -0.02,\ldots,-0.08 \) (from left to right). 

\newpage
\begin{center}
\epsfile{file=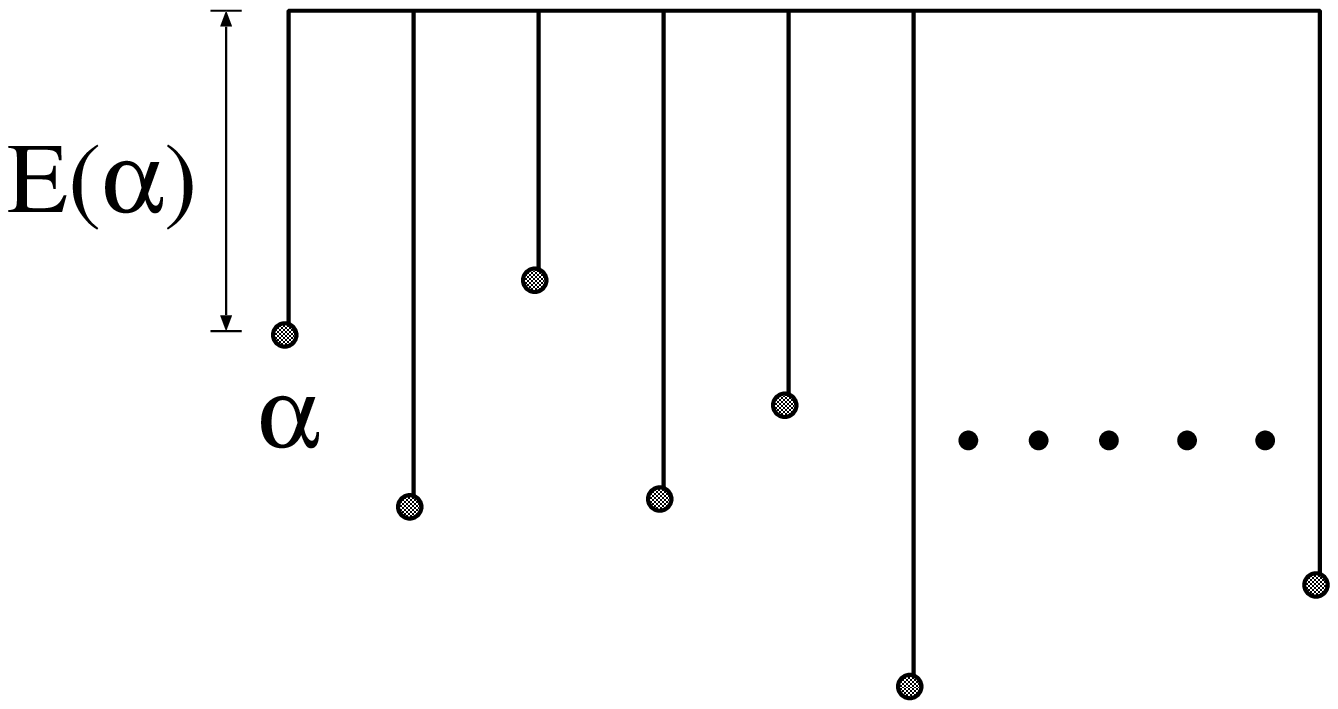,width=16.5cm}
\end{center}
\vspace{2cm}
\begin{center}
{\LARGE Fig.~1}
\end{center}
\newpage
\begin{center}
\epsfile{file=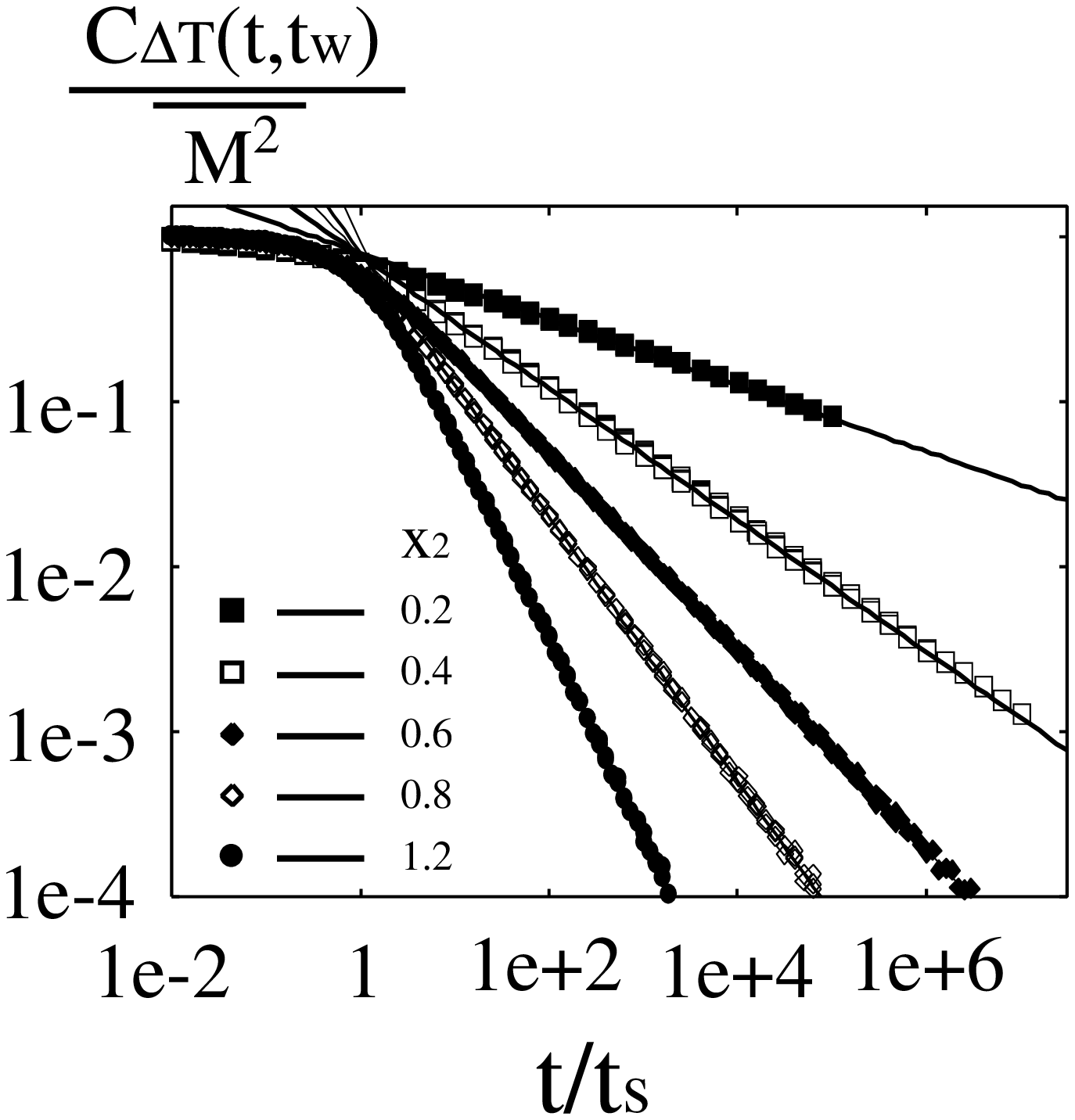,width=16.5cm}
\end{center}
\vspace{2cm}
\begin{center}
{\LARGE Fig.~2(a)}
\end{center}
\newpage
\begin{center}
\epsfile{file=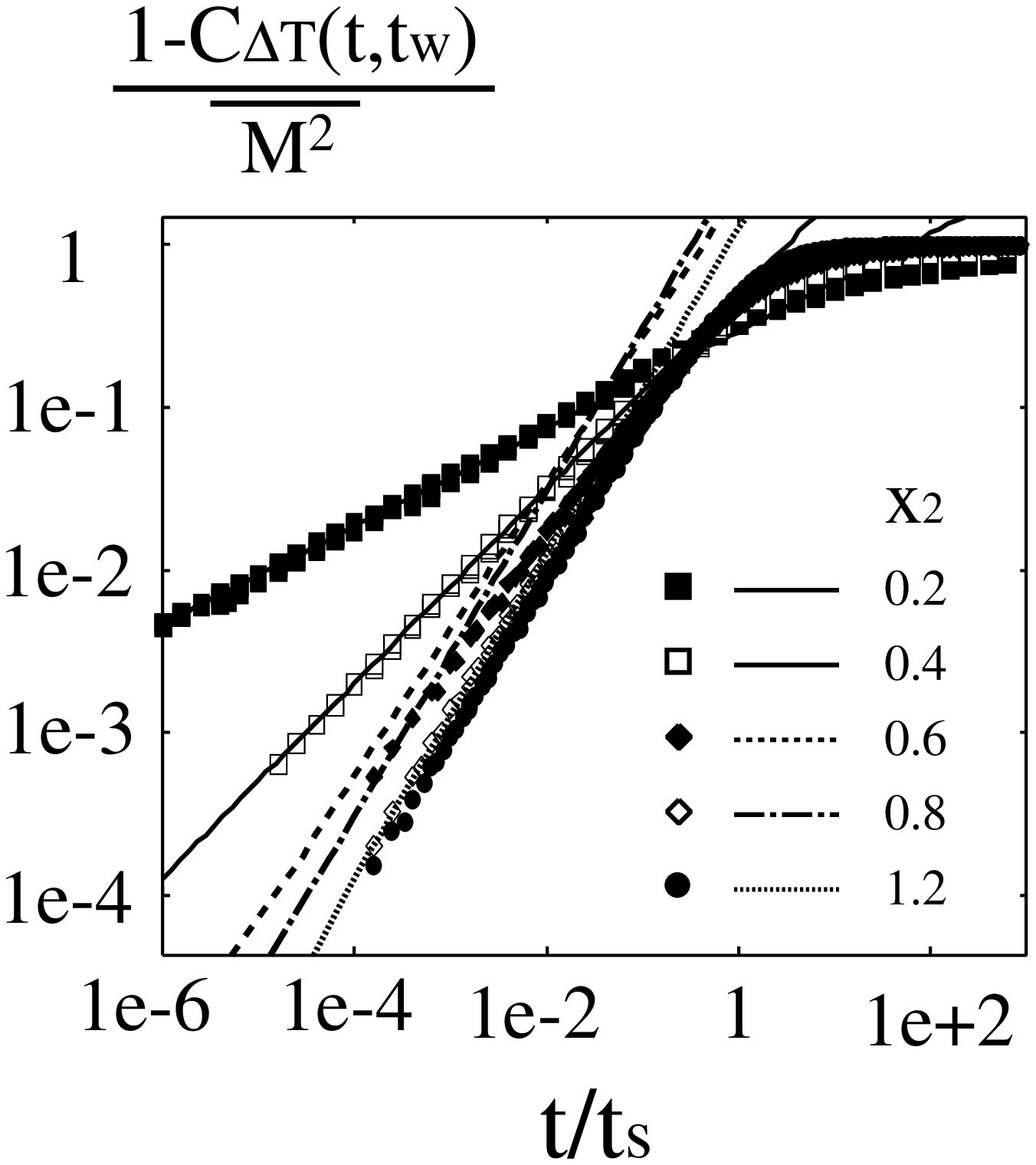,width=16.5cm}
\end{center}
\vspace{2cm}
\begin{center}
{\LARGE Fig.~2(b)}
\end{center}
\newpage
\begin{center}
\epsfile{file=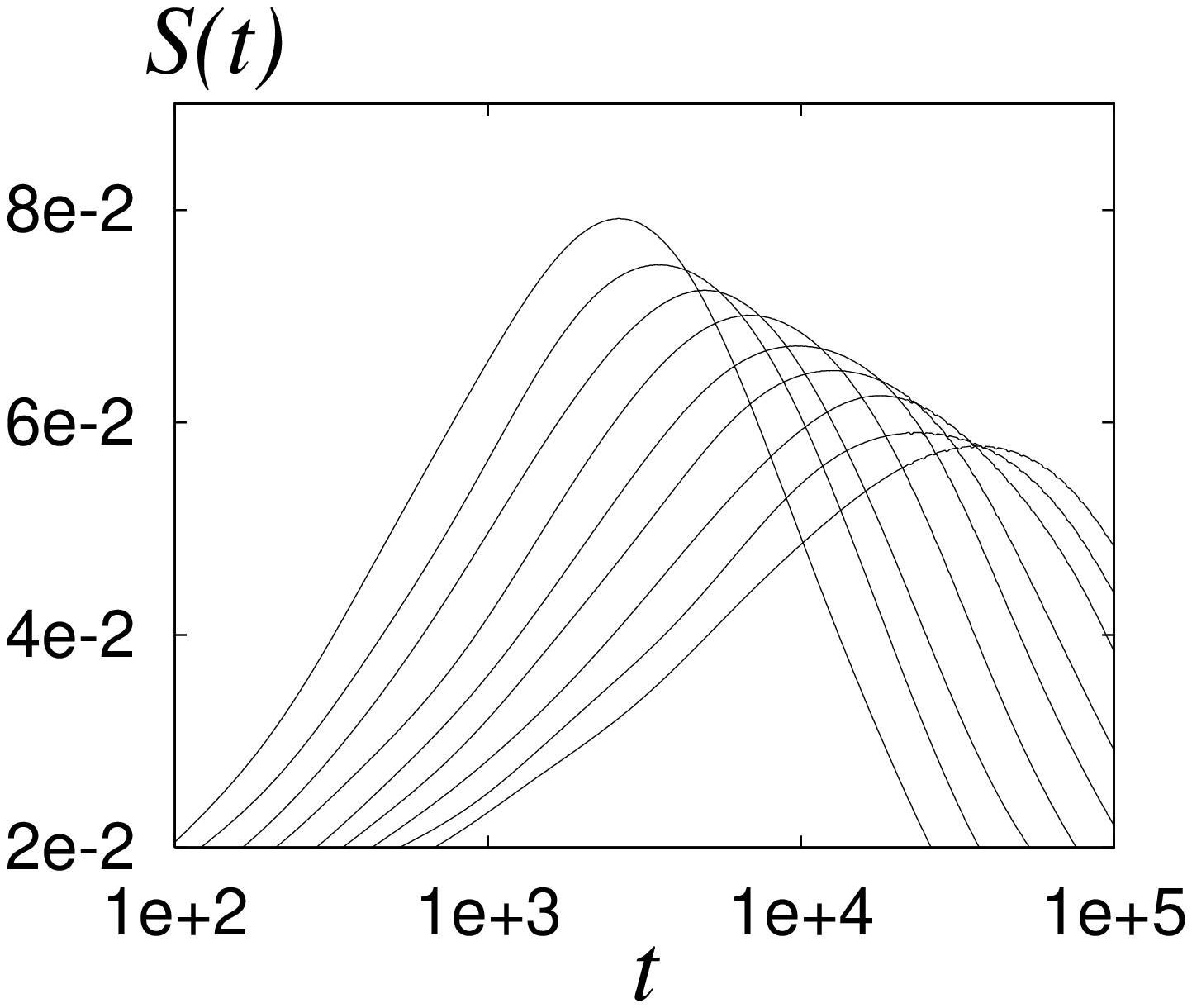,width=16.5cm}
\end{center}
\vspace{2cm}
\begin{center}
{\LARGE Fig.~3}
\end{center}
\newpage
\begin{center}
\epsfile{file=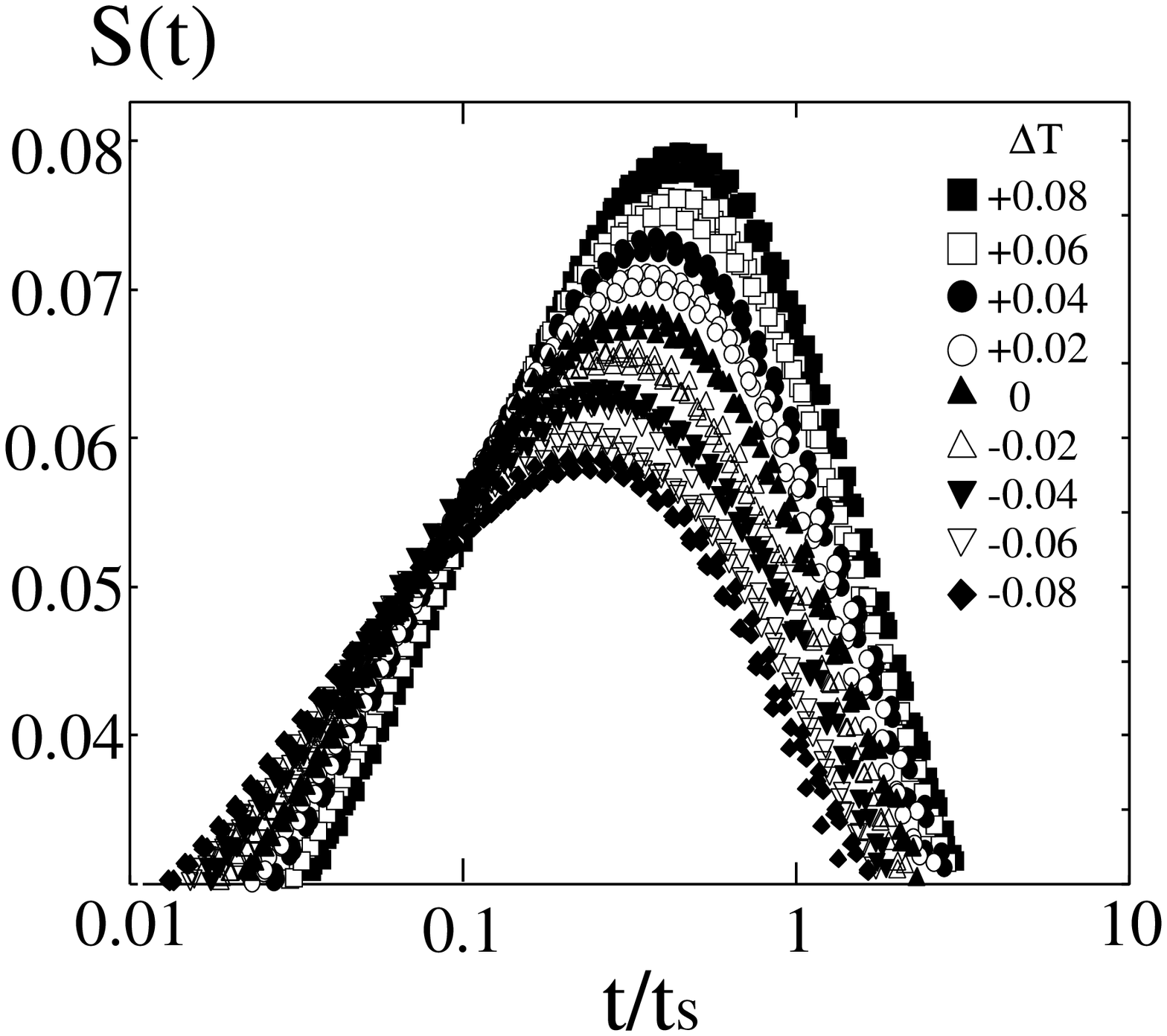,width=16.5cm}
\end{center}
\vspace{2cm}
\begin{center}
{\LARGE Fig.~4}
\end{center}
\newpage
\begin{center}
\epsfile{file=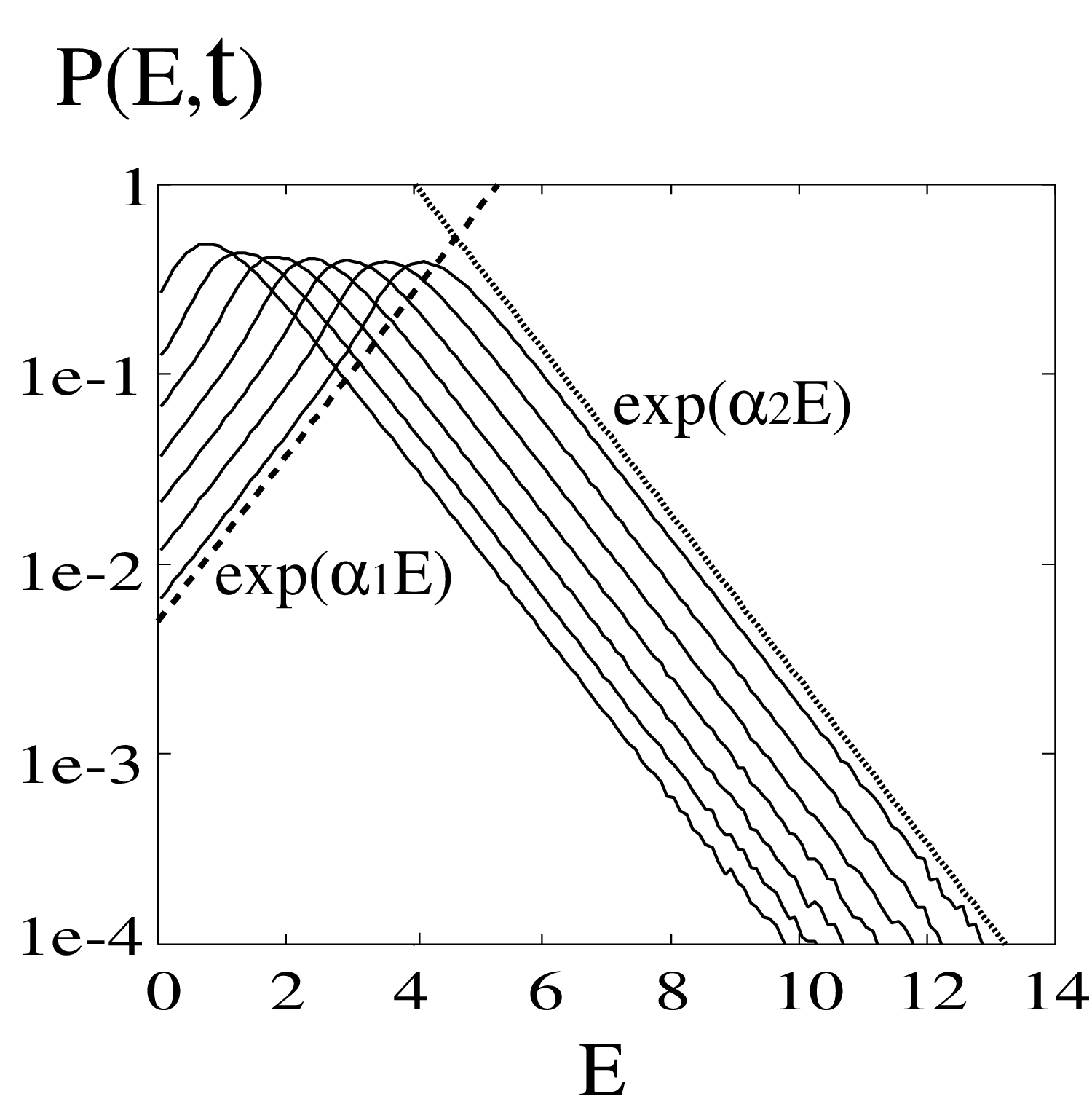,width=16.5cm}
\end{center}
\vspace{2cm}
\begin{center}
{\LARGE Fig.~5}
\end{center}
\newpage
\begin{center}
\epsfile{file=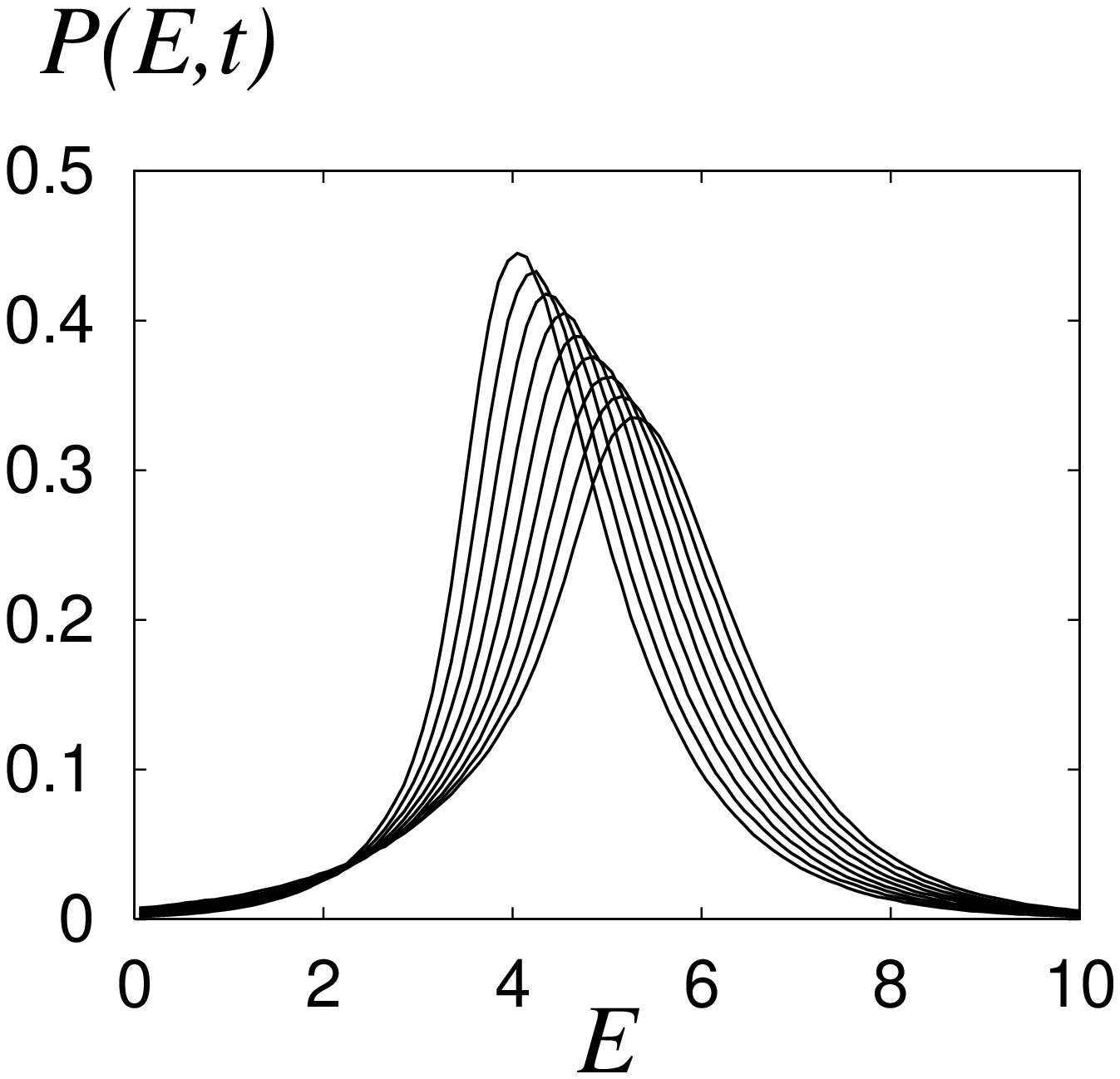,width=16.5cm}
\end{center}
\vspace{2cm}
\begin{center}
{\LARGE Fig.~6}
\end{center}
\end{document}